\documentclass{article}
\usepackage{spconf}
\usepackage{amsthm} 
\usepackage{dblfloatfix}
\theoremstyle{plain}

\usepackage{cite}
\usepackage{amsmath,amssymb,amsfonts}
\usepackage{algorithmic}
\usepackage{graphicx}
\usepackage{textcomp}
\usepackage{xcolor}
\usepackage{color,soul}

\usepackage{float}
\usepackage{placeins}
\usepackage{array}    % for *{…}{…} repetitions
\usepackage{siunitx}  % for the S column type
\usepackage{booktabs} % for \toprule, \midrule, \bottomrule
\usepackage{etoolbox}
\makeatletter
\apptocmd{\thebibliography}{%
  \setlength{\parskip}{0pt}%
  \setlength{\itemsep}{0pt}%
  \setlength{\parsep}{0pt}%
  \setlength{\topsep}{0pt}%
  \setlength{\partopsep}{0pt}%
}{}{}
\makeatother
\def\BibTeX{{\rm B\kern-.05em{\sc i\kern-.025em b}\kern-.08em
    T\kern-.1667em\lower.7ex\hbox{E}\kern-.125emX}}
\title{Ranking the Impact of Contextual Specialization in Neural Speech Enhancement}
\name{Peter Leer$^{\,\star,\dagger}$, Svend Feldt$^{\,\star}$, Zheng-Hua Tan$^{\,\dagger}$, Jan Østergaard$^{\,\dagger}$, Jesper Jensen$^{\, \star,\dagger}$ \thanks{This work is partly supported by Innovation Fund Denmark Case no. 4298-00010B}}
\address{$^{\star}$ Eriksholm Research Centre, Snekkersten, Denmark \\
$^{\dagger}$Aalborg University, Department of Electronic Systems, Aalborg, Denmark}
\usepackage{hyperref} % load last

\begin{document}
\ninept
\maketitle
\begin{abstract}
We systematically investigate neural speech enhancement systems, ranging from very small ($\sim$10\,k parameters) to medium-large ($\sim$2-5\,M parameters), which specialize to acoustic conditions using contextual information such as speaker identity, noise type, speaker gender, spoken language, and SNR. By fine-tuning generalist models on specific data subsets, we find that specializing to a speaker's identity consistently yields the largest gains in estimated speech intelligibility and quality. In contrast, specializing to SNR, noise type, or gender offers only marginal benefits. Crucially, we show that a small model specialized to both a specific speaker and a specific noise type can match or exceed the performance of a generalist model ten times its size. Further, cross-lingual tests reveal that models specialized to a target language outperform multilingual generalists, suggesting that language is a salient feature for specialization. These findings highlight the potential of small, adaptive models for resource-constrained applications like hearing aids, which specialize on-the-fly to contextual information.

\end{abstract}
\section{Introduction}

\label{sec:intro}
Understanding speech in noisy situations is a common challenge, especially for people with impaired hearing \cite{pichora-fuller_effects_2003}. To address this, edge devices, such as hearing aids, are often prescribed. However, despite substantial progress in hearing-aid technology and signal processing, enhancing speech intelligibility (SI) and speech quality (SQ) of noisy speech  in real-world scenarios (e.g., crowded restaurants and public transport) remains a significant challenge \cite{mustafa_comprehensive_2025}.

Neural network-based speech-enhancement models have demonstrated impressive improvements in both  SI and SQ \cite{hu_dccrn_2020,luo_conv-tasnet_2019}. Many high-performing models are trained on large, diverse datasets that allow the models to generalize across a wide range of speakers, noise types, and acoustic conditions \cite{reddy_interspeech_2020}. While these models (referred to as generalists in the remainder of this paper) can provide robust generalization performance in unseen settings, the robustness often comes at a significant memory and computational cost. As a result, these generalist models are typically too resource-intensive for deployment on embedded platforms such as hearing aids.

In many practical scenarios, an individual's acoustic environment is often predictable and relatively stationary. A particular individual often interacts with a small set of familiar voices - such as family members, coworkers, or caregivers \cite{plante-hebert_processing_2021} - in recurring acoustic environments like their homes, workplaces, or their transit routes. Similarly, the types of background noise encountered in these environments - such as traffic, kitchen noise, competing speakers - are relatively few and can be characterized and used to estimate model parameters \cite{lee_nastar_2022}. This stationary and predictable pattern suggests that models could benefit from adapting to these recurring conditions on-the-fly. If such adaptation enables small specialist models to deliver high performance in contextually familiar and stationary settings, it could dramatically reduce the resource requirements for effective speech enhancement (SE) on edge devices. These observations motivate a deeper investigation into the potential of smaller specialist SE systems over traditional generalist systems: (i) What can be gained by adapting SE systems to specific contextual information, and (ii) which types of contextual information offer the largest potential for performance gains?

Previous work \cite{kolbaek_speech_2017} compared generalist SE systems (DNNs trained on a large variety of speakers and noises) with specialist systems (DNNs trained on a subset of speakers, SNRs, or noise types) using a fully connected feed-forward architecture. The results showed that training to the target speaker or to a particular noise type boosted intelligibility \textit{when all other factors were held constant} (e.g., a speaker specialist was tested for a fixed noise type and SNR), while specializing by gender or input SNR offered no measurable benefit.
In contrast, \cite{sivaraman_sparse_2020, sivaraman_zero-shot_2021} proposed a lightweight personalization framework based on mixture-of-experts gating that activates a specialist from a discrete set of options, based on gender, input SNR, or other cues. It was found that small models can outperform larger generalist models when specialized to the speaker identity of the current speaker. However, these studies leave several questions unanswered. Their findings were based on a single model architecture (or scaled variants thereof), making it unclear if the observed benefits are general or an artifact of their specific model choice. Furthermore, no prior work has systematically compared a comprehensive set of contextual information to establish a clear hierarchy of importance. This leaves a gap in understanding: Which types of contextual information offers the most potential gain for specialization, and do these findings hold true across a diverse set of modern SE architectures?
  
To address these gaps, this paper provides a systematic analysis of speech enhancement model specialization. We limit our scope to single-channel enhancement systems under additive noise and employ a seen-speaker/seen-acoustic-scene design to specifically isolate the effects of contextual adaptation, distinct from generalization. Within this setup, our contributions are three-fold: (1) we establish the generality of our findings by testing across a diverse set of DNN architectures; (2) we systematically rank contextual factors (speaker, noise, gender, and SNR) by establishing an empirical performance upper bound; (3) we conduct the first controlled investigation into language as a specialization factor, demonstrating a statistically significant, but small, performance advantage for language-specific models.

\section{DNN Architectures}
We evaluate a set of network architectures, including feedforward, convolutional, recurrent, and attention-based designs: A classic fully-connected neural network (FFNN) \cite{gonzalez_assessing_2023} is implemented alongside Conv-TasNet \cite{luo_conv-tasnet_2019}, a fully convolutional time-domain separation model. We also include LiSenNet \cite{yan_LiSenNet_2024}, DCCRN \cite{hu_dccrn_2020}, and TF-GridNet \cite{wang_tf-gridnet_2023}, which all adopt an encoder–decoder structure, but differ in their specific encoder and decoder architectures as well as in the module applied to the encoder’s output: LiSenNet uses a GRU-based dual-path block, DCCRN uses a complex-valued LSTM, and TF-GridNet uses both an LSTM and a self-attention based approach. To study the impact of model capacity, we produce three scaled variants of both FFNN and LiSenNet: “tiny (T)” ($\sim10$\,k parameters), “small (S)” ($\sim100$\,k parameters), and “medium (M)” ($\sim1$\,M parameters). For FFNN, scaling is achieved by increasing the size of each hidden layer identically. For LiSenNet, we adjust the dimensionality of the embedding block, which in turn alters the parameter counts in its encoder, embedding module, and decoder. We denote networks as (\#networktype-\#size), e.g., "FFNN-T" for a tiny FFNN or "DCCRN" for the default DCCRN.

\section{Experiments}
We conducted two experiments to evaluate model specialization. Experiment 1 compares the performance of SE models specialized to speakers, gender, noise type, and SNR to generalists. Experiment 2 focuses on language specialization by comparing an English-only model to a multilingual generalist.

\subsection{Experiment 1: Speaker-, Gender-, Noise- and SNR-Specialists}
 We first introduce the procedure used to generate the noisy speech mixture dataset, followed by how each specialist is defined and linked to the mixture dataset, and finally, how the DNNs are trained.

\subsubsection{Clean speech recordings}
Clean speech is drawn from Clarity \cite{graetzer_dataset_2022} and VCTK \cite{veaux_voice_2013}. For each speaker, we assign 70\% of utterances to training, 15\% to validation, and 15\% to testing. 

\subsubsection{Noise recordings}
We use the first microphone channel from DEMAND \cite{thiemann_demand_2013} and ARTE \cite{weisser_ambisonic_2019}, totaling 31 different recordings of different acoustic environments lasting five minutes each. We split every recording temporally: the first 70 \% for training, the next 15 \% for validation, and the final 15 \% for testing. These splits result in non-overlapping segments from the same acoustic scene.

\subsubsection{Mixture generation}
Mixtures are created by pairing clean speech and noise drawn from the same split to prevent data leakage. For the generalist training dataset, we generate 100 hours of mixtures by randomly sampling speakers and noises from the training splits of the clean speech and noise recordings, with SNRs uniformly randomly sampled in the interval [-10,10]\,dB. The validation dataset is created in a similar manner but only 2 hours of mixtures are generated. We use a similar procedure when generating datasets for the specialists, but instead we create 10 hours of training mixtures for each specialist (see Sec. \ref{sec:specialists})  and 1 hour of validation mixtures from the validation splits. To create the test dataset, we sample speakers (5 male and 5 female)
 at random without replacement from each clean-speech corpus (20 speakers total). Then, for every noise type (31 in total), we create a 30\,s mixture for each randomly sampled speaker at each of the five SNRs $\{-10,-5,0,5,10\}$\,dB, yielding $20 \times 31 \times 5 = 3100$ unique test sets (25.8\,h in total). 
For all datasets, each mixture is rescaled to an absolute RMS level of $-30\,\mathrm{dBFS}$, avoiding clipping and ensuring normalized signals for fair comparisons. 

\subsubsection{Specialist definitions}
\label{sec:specialists}
Speaker specialists (\textbf{Spk}) are trained using only the training utterances of a single speaker, mixed with training segments from all noise recordings. The speaker specialists are evaluated on that single speaker’s test utterances mixed with test segments of all noises. Noise-type specialists (\textbf{Ns}) are trained using only the training segment of a single noise recording, mixed with training utterances from all speakers; they are evaluated on the test segments of that noise mixed with test utterances. 
 SNR-specialists (\textbf{SNR}) are trained with noisy mixtures at single fixed SNR and are evaluated on the subset of test mixtures created at that specific SNR. Gender-specialists (\textbf{Gdr}) are trained by splitting the clean-speaker datasets into male and female speakers and evaluated on the corresponding gendered partition of the test set. Joint speaker-noise specialists \textbf{(Spk+Ns)} are trained analogously to the speaker and noise-type specialists: for each (speaker,\,noise) pair, we train on that speaker’s training utterances mixed with the training portion of that noise recording and evaluate on the corresponding test partitions. We focus on speakers and noises because conditioning on either factor alone has been shown to be effective in previous work \cite{kolbaek_speech_2017,sivaraman_sparse_2020,tsangko_dfingernet_2025}; we deliberately avoid higher-order joint specialists (e.g., Spk+Ns+SNR) to prevent combinatorial growth and excessive computations.

\subsubsection{Training}
To train the generalist DNNs we reuse each architecture’s original training configuration from its source paper, i.e., loss function, optimizer, learning-rate schedule, weight decay, gradient clipping, and other hyperparameters.  The only deviation is for LiSenNet, where we remove the PESQ \cite{noauthor_perceptual_2001} term from the loss; in pilot runs this term increased training time tenfold without significantly improving performance. We train the generalist for at most 100 epochs, using the checkpoint that achieved the best validation score, as measured by the individual loss functions used.
To train the specialists, we adopt a fine-tuning strategy, which we found was more efficient than training each specialist from scratch. Each specialist is initialized from its corresponding generalist checkpoint and fine-tuned for at most 10 epochs, which in our experiments was found sufficient for convergence. We use the checkpoint that generated the best validation score. We resume the optimizer state from the generalist when fine-tuning.

\subsection{Experiment 2: Language Specialists}
We investigate whether a SE model specialized only on English target speech demonstrates a performance advantage for English speech compared to a multilingual generalist model. In general, isolating such a language-dependent effect is difficult as comparisons across different datasets often confound linguistic factors with variables like speaker characteristics, recording equipment, and acoustic environments. Using EMIME \cite{wester_emime_2010}, which provides Finnish (FIN) and German (GER) speakers producing both their native language (L1) and English (ENG) in the same room and recorded with the same equipment, we compare an English-only specialist (S) to a multilingual generalist (G), trained using an otherwise identical procedure, except with another set of clean speakers for training mixture generation \cite{conneau_fleurs_2022}.

For each model, $M \in \{S,G\}$, language, $L\!\in\!\{\text{ENG,L1}\}$, speaker, $p$, and matched scene, $c$ (noise type, SNR), we compute SE metrics $y_{M,L,p,c}$ and average within speaker:
\begin{equation} 
m_{M,L,p} \triangleq \frac{1}{|\mathcal{C}_p|}\sum_{c\in\mathcal{C}_p} y_{M,L,p,c} \, ,
\end{equation}
where $\mathcal{C}_p$ is the set of matched scenes. The test sets are generated identically to Experiment 1, with only a change in the clean-speaker identities (i.e., the FIN and GER subsets of EMIME). For each target speaker, a matched test set is created for each scene and SNR. 

A naive comparison, such as directly observing $m_{S,\mathrm{ENG},p} > m_{G,\mathrm{ENG},p}$, is insufficient, since any difference could stem from one model having a superior general capacity, independent of language. Instead, we can test if $m_{S,\mathrm{ENG},p} > m_{S,\mathrm{L1},p}$. By doing so, we can control for fixed speaker- and model-level biases. To do so, define
$ \Delta_S \triangleq m_{S,\mathrm{ENG},p}-m_{S,\mathrm{L1},p}$. Then, an appropriate statistical test would be to test if $\Delta_S$ is non-zero.  
However, $\Delta_S$ is not an unbiased measure of specialization; it also captures confounding effects such as inherent metric biases or baseline differences in task difficulty between languages. Hence, to isolate the specialization effect, we finally use the multilingual generalist's performance difference as a contrast and estimate a per-speaker Model$\times$Language interaction:
\begin{align}
\delta_p &\triangleq \Delta_S - \Delta_G= \big(m_{S,\mathrm{ENG},p}-m_{S,\mathrm{L1},p}\big) - \big(m_{G,\mathrm{ENG},p}-m_{G,\mathrm{L1},p}\big) \nonumber \\
&= \big(m_{S,\mathrm{ENG},p}-m_{G,\mathrm{ENG},p}\big) - \big(m_{S,\mathrm{L1},p}-m_{G,\mathrm{L1},p}\big) \, ,
\label{eq:delta_p}
\end{align} 
such that a positive $\delta_p$ implies the Specialist–Generalist contrast is larger in English than in L1 for speaker $p$. 
\section{Results}

In this Section, we present and discuss results from the two experiments. We report SI-SDR \cite{roux_sdr_2018}, PESQ \cite{noauthor_perceptual_2001} and ESTOI \cite{jensen_algorithm_2016}
    \subsection{Experiment 1: Speaker-,Gender-,Noise-, and SNR-specialists }
     The average performance improvements over the unprocessed noisy speech, for all models and specialization configurations, are shown in Table \ref{tab:leaveoneout-results}. To validate our results, we performed a statistical analysis using pairwise Wilcoxon signed-rank tests with Holm-Bonferroni correction (adj. $p < 0.05$), where we performed comparisons within each model  (i.e., $6 \choose{2}$ $=15$ comparisons per architecture) across all test sets. 
    
    The analysis confirms that specialization yields significant gains across all metrics. Speaker-specialized models (Spk and Spk+Ns) consistently provide the largest improvements over the generalist (G), with Spk+Ns as the top-performing configuration, followed by Spk. Thus, we establish the rank:  Spk + Ns $>$ Spk $>$ SNR $\approx$ Ns $\approx$ Gdr $>$ G in terms of performance for all metrics. Within SNR/Ns/Gdr the exact ranking varies by architecture and the performance differences are relatively small. Note that the ranking holds across architectures except for ConvTasNet, where gender specialization reduced performance. Ablation experiments show that this performance reduction is due to the high learning rate used in the original setup; lowering it restores the ordering. We report the original setting in Table \ref{tab:leaveoneout-results} for consistency.
    
    Interestingly, specialization proves to outperform a model that is 10 times larger: We found that a smaller model with Spk+Ns specialization significantly outperforms a 10x larger generalist counterpart for all metrics and models, except for the ESTOI measurement of FFNN-T Spk+Ns vs. FFNN-S Base where they were not significantly different (i.e., performed on par). 

    We further analyzed whether the performance gains from speaker (Spk) and noise (Ns) specialization could be additively combined to predict the performance of the joint Spk+Ns model. The additive relationship was strong: On average, the predicted improvement for SI-SDR, PESQ and ESTOI differed from the actual improvement by only -0.04 dB, -0.007, and +0.001, respectively. The Conv-TasNet architecture was an exception, where the joint optimization outperformed the individual contributions significantly.

\begin{figure*}[!t]
    \centering
    \includegraphics[width=\textwidth]{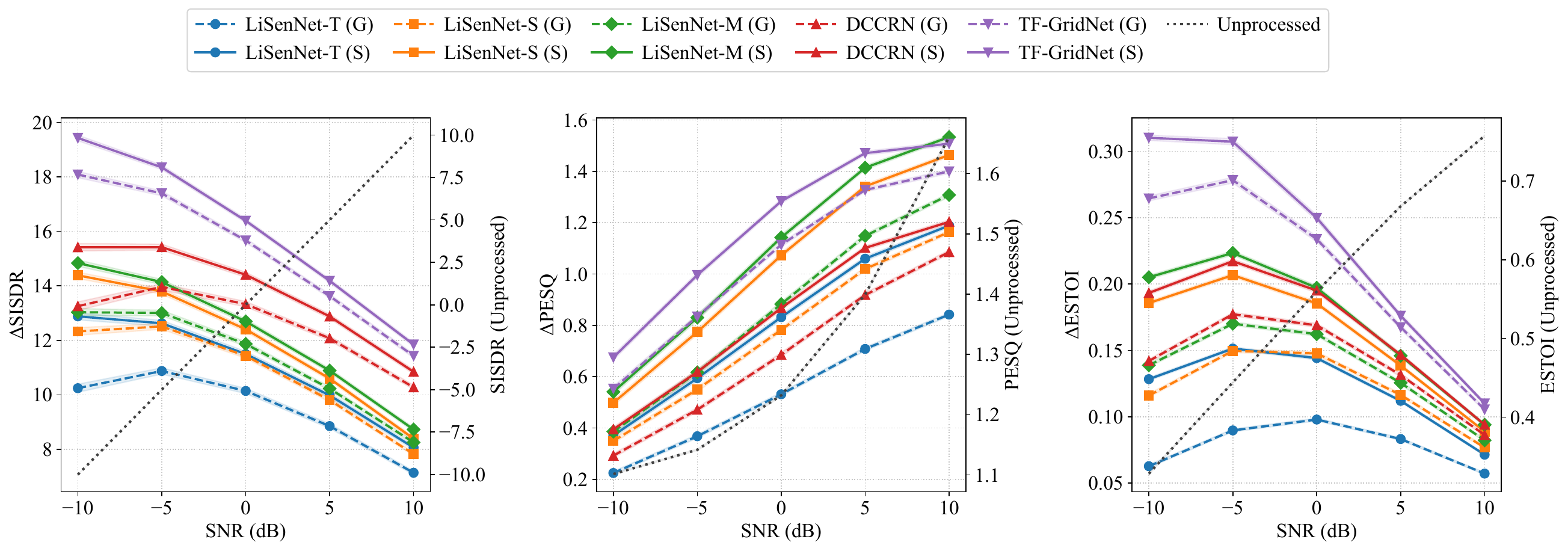}
    \caption{Experiment 1: Comparison of performance between selected generalist networks (G) and joint speaker-and-noise specialists (S). $\Delta$ denotes improvement from the unprocessed baseline.}
    \label{fig:snr_subplots}
\end{figure*}

In Figure \ref{fig:snr_subplots}, we compare the performance of a subset of the generalist and 'Spk+Ns' specialist networks as a function of SNR. Notably, the difference between generalist and specialist models is most pronounced at lower SNRs, suggesting that specialization provides the greatest benefit when the enhancement task is most difficult.

\begin{table}[h]
\centering
\caption{Experiment 1: Average improvement $(\Delta)$ in different metrics relative to the unprocessed mixture (Noisy scores are SI-SDR = -0.16 dB, PESQ = 1.31, ESTOI = 0.551). ‘G’ denotes the generalist; ‘SNR’, ‘Gdr’, ‘Spk’, and ‘Ns’ denote specialists for SNR, gender, speaker, and noise type; ‘Spk+Ns’ is joint speaker–noise specialization.}
\label{tab:leaveoneout-results}
\setlength{\tabcolsep}{3pt}
\footnotesize
\sisetup{table-number-alignment=center}

\textit{(a) $\Delta$SI-SDR [dB] (Noisy = -0.16).}
\begin{tabular}{l*{6}{S[table-format=2.2]}}
\toprule
\textbf{Architecture} & {G} & {SNR} & {Gdr} & {Spk} & {Ns} & {Spk+Ns} \\
\midrule
FFNN-T     &  6.59 &  6.96 &  7.09 &  7.88 &  7.56 &  8.61 \\
FFNN-S     &  8.50 &  8.79 &  8.82 &  9.45 &  8.93 &  9.73 \\
FFNN-M     &  8.99 &  9.17 &  9.21 &  9.80 &  9.19 &  9.95 \\
LiSenNet-T &  9.45 &  9.57 &  9.71 & 10.50 &  9.93 & 11.02 \\
LiSenNet-S & 10.80 & 11.07 & 11.00 & 11.57 & 11.10 & 11.94 \\
LiSenNet-M & 11.29 & 11.62 & 11.46 & 12.04 & 11.58 & 12.28 \\
DCCRN       & 12.58 & 12.76 & 12.79 & 13.73 & 12.78 & 13.80 \\
Conv-TasNet  & 14.50 & 14.61 & 13.04 & 15.29 & 14.47 & 15.50 \\
TF-GridNet   & 15.26 & 15.37 & 15.36 & 15.97 & 15.41 & 16.07 \\
\bottomrule
\end{tabular}

\vspace{2mm}

\textit{(b) $\Delta$PESQ (Noisy = 1.31).}
\begin{tabular}{l*{6}{S[table-format=1.2]}}
\toprule
\textbf{Architecture} & {G} & {SNR} & {Gdr} & {Spk} & {Ns} & {Spk+Ns} \\
\midrule
FFNN-T     & 0.21 & 0.23 & 0.25 & 0.29 & 0.29 & 0.35 \\
FFNN-S     & 0.35 & 0.36 & 0.38 & 0.45 & 0.40 & 0.49 \\
FFNN-M     & 0.40 & 0.41 & 0.42 & 0.50 & 0.43 & 0.52 \\
LiSenNet-T & 0.54 & 0.58 & 0.59 & 0.72 & 0.64 & 0.81 \\
LiSenNet-S & 0.78 & 0.81 & 0.82 & 0.97 & 0.85 & 1.03 \\
LiSenNet-M & 0.87 & 0.91 & 0.90 & 1.07 & 0.91 & 1.10 \\
DCCRN       & 0.69 & 0.72 & 0.71 & 0.82 & 0.74 & 0.84 \\
Conv-TasNet  & 0.79 & 0.79 & 0.49 & 0.88 & 0.78 & 0.91 \\
TF-GridNet   & 1.05 & 1.06 & 1.07 & 1.17 & 1.07 & 1.19 \\
\bottomrule
\end{tabular}

\vspace{2mm}

\textit{(c) $\Delta$ESTOI (Noisy = 0.551).}
\begin{tabular}{l*{6}{S[table-format=1.3]}}
\toprule
\textbf{Architecture} & {G} & {SNR} & {Gdr} & {Spk} & {Ns} & {Spk+Ns} \\
\midrule
FFNN-T     & 0.028 & 0.031 & 0.032 & 0.048 & 0.041 & 0.062 \\
FFNN-S     & 0.063 & 0.066 & 0.068 & 0.088 & 0.073 & 0.098 \\
FFNN-M     & 0.078 & 0.082 & 0.084 & 0.106 & 0.085 & 0.111 \\
LiSenNet-T & 0.078 & 0.079 & 0.083 & 0.101 & 0.087 & 0.121 \\
LiSenNet-S & 0.121 & 0.127 & 0.129 & 0.155 & 0.128 & 0.161 \\
LiSenNet-M & 0.136 & 0.145 & 0.144 & 0.170 & 0.145 & 0.173 \\
DCCRN       & 0.141 & 0.144 & 0.146 & 0.166 & 0.146 & 0.169 \\
Conv-TasNet  & 0.168 & 0.169 & 0.132 & 0.182 & 0.169 & 0.189 \\
TF-GridNet   & 0.210 & 0.212 & 0.212 & 0.229 & 0.212 & 0.231 \\
\bottomrule
\end{tabular}
\end{table}
\begin{table}[!htb]
\centering
\caption{Experiment 2, $\delta_p$ (see Eq. ~\ref{eq:delta_p}) pooled per speaker across SNRs. A positive value means the model trained on only English has a larger English advantage than the multilingual model. Boldface indicates Benjamini-Hochberg-corrected significant at $q < 0.05$.}
\label{tab:emime-dod-fdr}
\sisetup{detect-weight=true,detect-inline-weight=math}

% Top part of the table: SI-SDR and PESQ
\begin{tabular}{l S[table-format=-1.3] S[table-format=1.3] S[table-format=-1.3] S[table-format=1.3]}
\toprule
\multicolumn{1}{c}{\textbf{Architecture}} &  \multicolumn{2}{c}{\textit{SI-SDR}} & \multicolumn{2}{c}{\textit{PESQ}} \\
\cmidrule(lr){2-3} \cmidrule(lr){4-5}
 & {FIN} & {GER} & {FIN} & {GER} \\
\midrule
FFNN-T     & \bfseries 0.142 & 0.015 & 0.000 & \bfseries 0.003 \\
FFNN-S     & \bfseries 0.115 & 0.005 & \bfseries 0.006 & \bfseries 0.005 \\
FFNN-M     & \bfseries 0.105 & 0.019 & \bfseries 0.010 & \bfseries 0.005 \\
LiSenNet-T & \bfseries 0.254 & 0.050 & \bfseries 0.029 & \bfseries 0.017 \\
LiSenNet-S & \bfseries 0.243 & 0.073 & \bfseries 0.053 & \bfseries 0.021 \\
LiSenNet-M & \bfseries 0.244 & \bfseries 0.068 & \bfseries 0.052 & \bfseries 0.018 \\
Conv-TasNet & -0.014 & 0.062 & -0.004 & \bfseries 0.021 \\
TF-GridNet  & 0.054 & 0.056 & \bfseries 0.050 & \bfseries 0.027 \\
\bottomrule
\end{tabular}

\vspace{2mm} % Space between the two stacked tables

% Bottom part of the table: ESTOI
\begin{tabular}{l S[table-format=1.3] S[table-format=1.3]}
\toprule
\multicolumn{1}{c}{\textbf{Architecture}} & \multicolumn{2}{c}{\textit{ESTOI}} \\
\cmidrule(lr){2-3}
 & {FIN} & {GER} \\
\midrule
FFNN-T     & \bfseries 0.004 & \bfseries 0.001 \\
FFNN-S     & \bfseries 0.006 & \bfseries 0.002 \\
FFNN-M     & \bfseries 0.006 & \bfseries 0.002 \\
LiSenNet-T & \bfseries 0.016 & \bfseries 0.004 \\
LiSenNet-S & \bfseries 0.017 & \bfseries 0.006 \\
LiSenNet-M & \bfseries 0.018 & \bfseries 0.005 \\
Conv-TasNet & \bfseries 0.004 & \bfseries 0.003 \\
TF-GridNet  & \bfseries 0.011 & \bfseries 0.004 \\
\bottomrule
\end{tabular}
\end{table}

\subsection{Experiment 2}
Table \ref{tab:emime-dod-fdr} shows performance improvement of English as compared to native languages (Finnish, German). 
For each model family and group (FIN/GER),  we perform a paired Wilcoxon signed-rank test across speakers. We control multiplicity within each metric using a Benjamini-Hochberg (BH) correction procedure.
 Our findings show a clear and consistent trend: The $\delta$ values are  positive and statistically significant across nearly all metrics and models. This indicates that the models trained exclusively on English data exhibit a performance advantage for English speech compared to the models trained on the multilingual FLEURS dataset. Absolute improvements are, however, modest.

Interestingly, this English advantage is consistently larger for Finnish speakers than for German speakers. This observation might be explained by the linguistic distance between the languages. English, being a Germanic language, is typologically and prosodically much closer to German than it is to Finnish.

\section{Conclusion}\label{sec:Conclusion}
We systematically evaluated the benefits of specializing speech enhancement models, framing the analysis as an empirical upper-bound performance scenario with oracle contextual information. This aligns with findings from \cite{sivaraman_sparse_2020, sivaraman_zero-shot_2021} and clarifies earlier work.
Our results show that speaker identity is the most valuable information for specialization, amongst the set of contextual information tested. Furthermore, a joint speaker-noise specialist performs best, with the gains from each specialization proving to be nearly additive.  Notably, we demonstrated that small specialized models can outperform generalist models over 10 times their size, making a strong case for adaptable systems on resource-constrained hardware. We also provided evidence suggesting that language specialization offers modest but consistent benefits.

\FloatBarrier
\newpage
%\clearpage

\bibliographystyle{IEEEbib}
\bibliography{strings,refs}

@misc{tsangko_dfingernet_2025,
	title = {{DFingerNet}: {Noise}-{Adaptive} {Speech} {Enhancement} for {Hearing} {Aids}},
	copyright = {arXiv.org perpetual, non-exclusive license},
	shorttitle = {{DFingerNet}},
	url = {https://arxiv.org/abs/2501.10525},
	doi = {10.48550/ARXIV.2501.10525},
	urldate = {2025-09-02},
	publisher = {arXiv},
	author = {Tsangko, Iosif and Triantafyllopoulos, Andreas and Müller, Michael and Schröter, Hendrik and Schuller, Björn W.},
	year = {2025},
	note = {Version Number: 2},
}

@misc{conneau_fleurs_2022,
	title = {{FLEURS}: {Few}-shot {Learning} {Evaluation} of {Universal} {Representations} of {Speech}},
	copyright = {Creative Commons Attribution 4.0 International},
	shorttitle = {{FLEURS}},
	url = {https://arxiv.org/abs/2205.12446},
	doi = {10.48550/ARXIV.2205.12446},
	urldate = {2025-08-31},
	publisher = {arXiv},
	author = {Conneau, Alexis and Ma, Min and Khanuja, Simran and Zhang, Yu and Axelrod, Vera and Dalmia, Siddharth and Riesa, Jason and Rivera, Clara and Bapna, Ankur},
	year = {2022},
	note = {Version Number: 1},
}

@misc{noauthor_perceptual_2001,
	title = {Perceptual evaluation of speech quality ({PESQ}): {An} objective method for end-to-end speech quality assessment of narrow-band telephone networks and speech codecs},
	url = {https://handle.itu.int/11.1002/1000/5374},
	publisher = {International Telecommunication Union (ITU-T), Study Group 12},
	month = feb,
	year = {2001},
}

@misc{roux_sdr_2018,
	title = {{SDR} - half-baked or well done?},
	copyright = {arXiv.org perpetual, non-exclusive license},
	url = {https://arxiv.org/abs/1811.02508},
	doi = {10.48550/ARXIV.1811.02508},
	urldate = {2025-08-28},
	publisher = {arXiv},
	author = {Roux, Jonathan Le and Wisdom, Scott and Erdogan, Hakan and Hershey, John R.},
	year = {2018},
	note = {Version Number: 1},
}

@article{jensen_algorithm_2016,
	title = {An {Algorithm} for {Predicting} the {Intelligibility} of {Speech} {Masked} by {Modulated} {Noise} {Maskers}},
	volume = {24},
	copyright = {https://ieeexplore.ieee.org/Xplorehelp/downloads/license-information/IEEE.html},
	issn = {2329-9290, 2329-9304},
	url = {https://ieeexplore.ieee.org/document/7539284/},
	doi = {10.1109/TASLP.2016.2585878},
	number = {11},
	urldate = {2025-08-28},
	journal = {IEEE/ACM Transactions on Audio, Speech, and Language Processing},
	author = {Jensen, Jesper and Taal, Cees H.},
	month = nov,
	year = {2016},
	pages = {2009--2022},
}

@misc{wester_emime_2010,
	title = {The {EMIME} {Bilingual} {Database}},
	author = {Wester, Mirjam},
	year = {2010},
}

@inproceedings{veaux_voice_2013,
	address = {Gurgaon, India},
	title = {The voice bank corpus: {Design}, collection and data analysis of a large regional accent speech database},
	isbn = {978-1-4799-2378-6},
	shorttitle = {The voice bank corpus},
	url = {http://ieeexplore.ieee.org/document/6709856/},
	doi = {10.1109/ICSDA.2013.6709856},
	urldate = {2025-08-06},
	booktitle = {2013 {International} {Conference} {Oriental} {COCOSDA} held jointly with 2013 {Conference} on {Asian} {Spoken} {Language} {Research} and {Evaluation} ({O}-{COCOSDA}/{CASLRE})},
	publisher = {IEEE},
	author = {Veaux, Christophe and Yamagishi, Junichi and King, Simon},
	month = nov,
	year = {2013},
	pages = {1--4},
}

@article{graetzer_dataset_2022,
	title = {Dataset of {British} {English} speech recordings for psychoacoustics and speech processing research: {The} clarity speech corpus},
	volume = {41},
	issn = {23523409},
	shorttitle = {Dataset of {British} {English} speech recordings for psychoacoustics and speech processing research},
	url = {https://linkinghub.elsevier.com/retrieve/pii/S2352340922001627},
	doi = {10.1016/j.dib.2022.107951},
	language = {en},
	urldate = {2025-08-06},
	journal = {Data in Brief},
	author = {Graetzer, Simone and Akeroyd, Michael A. and Barker, Jon and Cox, Trevor J. and Culling, John F. and Naylor, Graham and Porter, Eszter and Viveros-Muñoz, Rhoddy},
	month = apr,
	year = {2022},
	pages = {107951},
}

@article{weisser_ambisonic_2019,
	title = {The {Ambisonic} {Recordings} of {Typical} {Environments} ({ARTE}) {Database}},
	volume = {105},
	issn = {1610-1928},
	url = {https://dael.euracoustics.org/landing_pages/aaua/52700.html},
	doi = {10.3813/AAA.919349},
	language = {en},
	number = {4},
	urldate = {2025-08-06},
	journal = {Acta Acustica united with Acustica},
	author = {Weisser, Adam and Buchholz, Jörg M. and Oreinos, Chris and Badajoz-Davila, Javier and Galloway, James and Beechey, Timothy and Keidser, Gitte},
	month = jul,
	year = {2019},
	pages = {695--713},
}

@misc{thiemann_demand_2013,
	title = {Demand: {A} {Collection} {Of} {Multi}-{Channel} {Recordings} {Of} {Acoustic} {Noise} {In} {Diverse} {Environments}},
	copyright = {Creative Commons Attribution 4.0, Open Access},
	shorttitle = {Demand},
	url = {https://zenodo.org/record/1227121},
	doi = {10.5281/ZENODO.1227121},
	language = {en},
	urldate = {2025-08-06},
	publisher = {Zenodo},
	author = {Thiemann, Joachim and Ito, Nobutaka and Vincent, Emmanuel},
	month = jun,
	year = {2013},
}

@misc{wang_tf-gridnet_2023,
	title = {{TF}-{GridNet}: {Making} {Time}-{Frequency} {Domain} {Models} {Great} {Again} for {Monaural} {Speaker} {Separation}},
	shorttitle = {{TF}-{GridNet}},
	url = {http://arxiv.org/abs/2209.03952},
	doi = {10.48550/arXiv.2209.03952},
	urldate = {2025-06-11},
	publisher = {arXiv},
	author = {Wang, Zhong-Qiu and Cornell, Samuele and Choi, Shukjae and Lee, Younglo and Kim, Byeong-Yeol and Watanabe, Shinji},
	month = mar,
	year = {2023},
	note = {arXiv:2209.03952 [cs]},
}

@misc{yan_lisennet_2024,
	title = {{LiSenNet}: {Lightweight} {Sub}-band and {Dual}-{Path} {Modeling} for {Real}-{Time} {Speech} {Enhancement}},
	shorttitle = {{LiSenNet}},
	url = {http://arxiv.org/abs/2409.13285},
	doi = {10.48550/arXiv.2409.13285},
	urldate = {2025-06-11},
	publisher = {arXiv},
	author = {Yan, Haoyin and Zhang, Jie and Fan, Cunhang and Zhou, Yeping and Liu, Peiqi},
	month = sep,
	year = {2024},
	note = {arXiv:2409.13285 [eess]},
}

@article{gonzalez_assessing_2023,
	title = {Assessing the {Generalization} {Gap} of {Learning}-{Based} {Speech} {Enhancement} {Systems} in {Noisy} and {Reverberant} {Environments}},
	volume = {31},
	issn = {2329-9290, 2329-9304},
	url = {http://arxiv.org/abs/2309.06183},
	doi = {10.1109/TASLP.2023.3318965},
	urldate = {2025-06-11},
	journal = {IEEE/ACM Transactions on Audio, Speech, and Language Processing},
	author = {Gonzalez, Philippe and Alstrøm, Tommy Sonne and May, Tobias},
	year = {2023},
	note = {arXiv:2309.06183 [eess]},
	pages = {3390--3403},
}

@misc{lee_nastar_2022,
	title = {{NASTAR}: {Noise} {Adaptive} {Speech} {Enhancement} with {Target}-{Conditional} {Resampling}},
	shorttitle = {{NASTAR}},
	url = {http://arxiv.org/abs/2206.09058},
	doi = {10.48550/arXiv.2206.09058},
	urldate = {2025-06-11},
	publisher = {arXiv},
	author = {Lee, Chi-Chang and Hu, Cheng-Hung and Lin, Yu-Chen and Chen, Chu-Song and Wang, Hsin-Min and Tsao, Yu},
	month = jun,
	year = {2022},
	note = {arXiv:2206.09058 [eess]},
}

@article{plante-hebert_processing_2021,
	title = {The processing of intimately familiar and unfamiliar voices: {Specific} neural responses of speaker recognition and identification},
	volume = {16},
	issn = {1932-6203},
	shorttitle = {The processing of intimately familiar and unfamiliar voices},
	url = {https://dx.plos.org/10.1371/journal.pone.0250214},
	doi = {10.1371/journal.pone.0250214},
	language = {en},
	number = {4},
	urldate = {2025-06-11},
	journal = {PLOS ONE},
	author = {Plante-Hébert, Julien and Boucher, Victor J. and Jemel, Boutheina},
	editor = {Fu, Qian-Jie},
	month = apr,
	year = {2021},
	pages = {e0250214},
}

@misc{hu_dccrn_2020,
	title = {{DCCRN}: {Deep} {Complex} {Convolution} {Recurrent} {Network} for {Phase}-{Aware} {Speech} {Enhancement}},
	shorttitle = {{DCCRN}},
	url = {http://arxiv.org/abs/2008.00264},
	doi = {10.48550/arXiv.2008.00264},
	urldate = {2025-06-11},
	publisher = {arXiv},
	author = {Hu, Yanxin and Liu, Yun and Lv, Shubo and Xing, Mengtao and Zhang, Shimin and Fu, Yihui and Wu, Jian and Zhang, Bihong and Xie, Lei},
	month = sep,
	year = {2020},
	note = {arXiv:2008.00264 [eess]},
}

@article{luo_conv-tasnet_2019,
	title = {Conv-{TasNet}: {Surpassing} {Ideal} {Time}-{Frequency} {Magnitude} {Masking} for {Speech} {Separation}},
	volume = {27},
	issn = {2329-9290, 2329-9304},
	shorttitle = {Conv-{TasNet}},
	url = {http://arxiv.org/abs/1809.07454},
	doi = {10.1109/TASLP.2019.2915167},
	number = {8},
	urldate = {2025-06-11},
	journal = {IEEE/ACM Transactions on Audio, Speech, and Language Processing},
	author = {Luo, Yi and Mesgarani, Nima},
	month = aug,
	year = {2019},
	note = {arXiv:1809.07454 [cs]},
	pages = {1256--1266},
}

@misc{reddy_interspeech_2020,
	title = {The {INTERSPEECH} 2020 {Deep} {Noise} {Suppression} {Challenge}: {Datasets}, {Subjective} {Testing} {Framework}, and {Challenge} {Results}},
	shorttitle = {The {INTERSPEECH} 2020 {Deep} {Noise} {Suppression} {Challenge}},
	url = {http://arxiv.org/abs/2005.13981},
	doi = {10.48550/arXiv.2005.13981},
	urldate = {2025-06-11},
	publisher = {arXiv},
	author = {Reddy, Chandan K. A. and Gopal, Vishak and Cutler, Ross and Beyrami, Ebrahim and Cheng, Roger and Dubey, Harishchandra and Matusevych, Sergiy and Aichner, Robert and Aazami, Ashkan and Braun, Sebastian and Rana, Puneet and Srinivasan, Sriram and Gehrke, Johannes},
	month = oct,
	year = {2020},
	note = {arXiv:2005.13981 [eess]},
}

@misc{sivaraman_zero-shot_2021,
	title = {Zero-{Shot} {Personalized} {Speech} {Enhancement} through {Speaker}-{Informed} {Model} {Selection}},
	url = {http://arxiv.org/abs/2105.03542},
	doi = {10.48550/arXiv.2105.03542},
	language = {en},
	urldate = {2025-06-11},
	publisher = {arXiv},
	author = {Sivaraman, Aswin and Kim, Minje},
	month = may,
	year = {2021},
	note = {arXiv:2105.03542 [eess]},
}

@article{mustafa_comprehensive_2025,
	title = {A comprehensive review on real-world challenges faced by hearing aid users and innovative solutions for background noise—from lab to life},
	volume = {41},
	issn = {2090-8539},
	url = {https://ejo.springeropen.com/articles/10.1186/s43163-025-00814-6},
	doi = {10.1186/s43163-025-00814-6},
	language = {en},
	number = {1},
	urldate = {2025-06-11},
	journal = {The Egyptian Journal of Otolaryngology},
	author = {Mustafa, Mishal and Krishnamurthy, Avinash},
	month = apr,
	year = {2025},
	pages = {62},
}

@article{pichora-fuller_effects_2003,
	title = {Effects of aging on auditory processing of speech},
	volume = {42 Suppl 2},
	issn = {1499-2027},
	language = {eng},
	journal = {International Journal of Audiology},
	author = {Pichora-Fuller, M. Kathleen and Souza, Pamela E.},
	month = jul,
	year = {2003},
	pmid = {12918623},
	pages = {2S11--16},
}

@misc{sivaraman_sparse_2020,
	title = {Sparse {Mixture} of {Local} {Experts} for {Efficient} {Speech} {Enhancement}},
	url = {http://arxiv.org/abs/2005.08128},
	doi = {10.48550/arXiv.2005.08128},
	urldate = {2025-05-01},
	publisher = {arXiv},
	author = {Sivaraman, Aswin and Kim, Minje},
	month = may,
	year = {2020},
	note = {arXiv:2005.08128 [eess]},
}

@article{kolbaek_speech_2017,
	title = {Speech {Intelligibility} {Potential} of {General} and {Specialized} {Deep} {Neural} {Network} {Based} {Speech} {Enhancement} {Systems}},
	volume = {25},
	copyright = {https://ieeexplore.ieee.org/Xplorehelp/downloads/license-information/IEEE.html},
	issn = {2329-9290, 2329-9304},
	url = {https://ieeexplore.ieee.org/document/7744475/},
	doi = {10.1109/taslp.2016.2628641},
	language = {en},
	number = {1},
	urldate = {2025-05-01},
	journal = {IEEE/ACM Transactions on Audio, Speech, and Language Processing},
	author = {Kolbæk, Morten and Tan, Zheng-Hua and Jensen, Jesper},
	month = jan,
	year = {2017},
	note = {Publisher: Institute of Electrical and Electronics Engineers (IEEE)},
	pages = {153--167},
}

%\bibliographystyle{IEEEbib}
%\bibliography{strings,refs}

\end{document}